\documentclass[12pt]{iopart}

\usepackage{bm}
\usepackage{graphicx}

\usepackage{latexsym}
\usepackage{amssymb}
\expandafter\let\csname equation*\endcsname\relax 
\expandafter\let\csname endequation*\endcsname\relax
\usepackage{amsmath}
\usepackage{xcolor}
\usepackage{cite}
\usepackage[normalem]{ulem}
\usepackage[colorinlistoftodos]{todonotes}

\def\be{\begin{equation}}
\def\ee{\end{equation}}
\def\bea{\begin{eqnarray}}
\def\eea{\end{eqnarray}}

\def\bi{\begin{itemize}}
\def\ei{\end{itemize}}
\def\ben{\begin{enumerate}}
\def\een{\end{enumerate}}

\usepackage{color}

\begin{document}

\title{Basis for time crystal phenomena in ultra-cold atoms bouncing on an oscillating mirror}

\author{Weronika Golletz$^1$, Andrzej Czarnecki$^2$, Krzysztof Sacha$^1$, Arkadiusz Kuro\'s$^{1,3}$}

\address{$^1$ Instytut Fizyki Teoretycznej, 
Uniwersytet Jagiello\'nski, ulica Profesora Stanis\l{}awa \L{}ojasiewicza 11, PL-30-348 Krak\'ow, Poland}

\address{$^2$ Instytut Matematyki, 
Uniwersytet Jagiello\'nski, ulica Profesora Stanis\l{}awa \L{}ojasiewicza 6, PL-30-348 Krak\'ow, Poland}

\address{$^3$ Institute of Physics, 
Jan Kochanowski University, ul. Uniwersytecka 7, 25-406 Kielce, Poland}
\ead{weronika.golletz@doctoral.uj.edu.pl}

\begin{abstract}
We consider classical dynamics of a 1D system of $N$ particles bouncing on an oscillating mirror in the presence of gravitational field. The particles behave like hard balls and they are resonantly driven by the mirror. We identify the manifolds the particles move on and derive the effective secular Hamiltonian for resonant motion of the particles. Proper choice of time periodic oscillations of the mirror allows for engineering of the effective behaviour of the particles. In particular, the system can behave like a $N$-dimensional fictitious particle moving in an $N$-dimensional crystalline structure.  Our classical analysis constitutes a basis for quantum research of novel time crystal phenomena in ultra-cold atoms bouncing on an oscillating atom mirror.
\end{abstract}
\noindent{\it Keywords\/}: periodically driven systems, secular Hamiltonians,  time crystals

\section{Introduction} \label{intro}

A particle moving in a one-dimensional (1D) space in the presence of gravitational field and bouncing on an oscillating mirror is one of the simplest Hamiltonian systems which can reveal a transition from regular to chaotic classical dynamics~\cite{Flatte1996,Buchleitner2002}. A simple criterion for the transition to chaos is related to the appearance of chaotic layers in the phase space of the particle when two neighbouring resonance islands start to overlap~\cite{Chirikov1979}. If, however, the amplitude of the mirror oscillations is not too large, the resonance islands are not destroyed and the phase space picture is regular. In quantum description there are states represented by localised wavepackets which show no spreading over time and evolve along resonant trajectories of classical particles~\cite{Buchleitner2002}. Recently, particles bouncing resonantly on an oscillating mirror became a playground for time crystal research~\cite{Sacha2015,Khemani16,ElseFTC,Sacha2017rev,khemani2019brief,Guo2020,SachaTC2020,GuoBook2021}. If  these particles are bosonic and the strength of the interactions between them is properly chosen, they can be made to spontaneously break the discrete time translation symmetry of the Hamiltonian and form discrete time crystals~\cite{Sacha2015,Giergiel2018a,Giergiel2018c,Kuros2020,Giergiel2020,Wang2020,Wang2021,Kuros2021}. Moreover, such a system is capable of realising analogues to various condensed matter phenomena in the time domain from Anderson localisation through topological time crystals, many-body localisation, Mott insulator phase to time lattices with exotic interactions~\cite{Sacha15a,Mierzejewski2017,Giergiel2018,Lustig2018,Giergiel2018b,Sharabi2021,SachaTC2020}. These time crystal systems can be realised in ultra-cold atom laboratories~\cite{Giergiel2018a,Giergiel2020}. Condensed matter can also be investigated in phase space crystals, see~\cite{Guo2013,Guo2016,Guo2016a,Liang2017,Guo2020,Guo2021,GuoBook2021}.

The case of a single particle bouncing between two orthogonal oscillating mirrors is reducible into two independent 1D motions along the perpendicular axes. However, for an arbitrary angle the system is no longer separable and the problem becomes more complex except for specific angles~\cite{wojtkowski1990,Whelan1990two,richter1990breathing, szeredi1993classical, szeredi1994classical, rouvinez1995classical}. For instance, let us consider the angle of $\pi/4$ with one of the mirrors remaining motionless along the gravitational field direction. Here the space available for particle's motion is halved with respect to the case of two orthogonal mirrors oscillating with the same period, phase, and amplitude~\cite{Giergiel2021}. This 2D motion can be mapped onto the problem of two particles in a 1D space interacting via a strong repulsive contact potential rendering the particles impenetrable (i.e. hard balls with zero radius)~\cite{wojtkowski1990}. 

In the present paper we consider $N$ impenetrable particles of equal masses stacked above one another in a 1D space bouncing on an oscillating mirror in the presence of gravitational field. We begin with the static mirror problem and analyse manifolds the impenetrable particles move on (\sref{static_mirror}). For $N=2$ and equal energies of the particles, the manifold they move on is the M\"obius strip~\cite{Giergiel2021}. 
We expand upon that model to treat the case  of $N=3$. We show that three particles with equal energies are moving on a solid torus. In \sref{two} we switch to the problem of an oscillating mirror and derive the effective secular Hamiltonian for impenetrable particles bouncing resonantly on the mirror. 

Our analysis is classical  in nature (i.e. it assumes classical distinguishable particles), but forms a basis for a fully quantum approach to investigations of novel quantum time crystals. In the quantum description $N$ classical particles are replaced by $N$ atomic clouds formed by different atomic species with atoms within each cloud being indistinguishable. The presence of intra-species interactions can lead to spontaneous breaking of discrete time translation symmetry and consequently, to the formation of various discrete time crystals. Apart from the spontaneous symmetry breaking phenomena, one can also investigate various condensed matter phases in the time domain because behaviour of the resonantly driven system can be reduced to solid state models. 
Note that if each of the $N$ clouds consists of only one atom, then due to the hard ball interactions, the system will possess properties of the Tonks–Girardeau gas~\cite{Girardeau1960}.

\section{Particles bouncing on a static mirror}
\label{static_mirror}

In this section, we deal with a 1D system of $N$ particles, stacked above one another, bouncing on a static mirror. First, we consider permeable particles using the standard action-angle variables $(I_i,\theta_i)$~\cite{Lichtenberg1992}. Next, we derive a general solution to a problem of $N$ impenetrable particles with different energies by introducing novel action-angle variables $({\cal I}_{i'},\vartheta_{i'})$. Then, we move to the case of equal energies. We focus on the systems of two and three impenetrable particles. The manifolds describing the motion for every case are analysed in detail.

\subsection{Action-angle variables for permeable particles} \label{secaa}
\label{aaper}

The Hamiltonian of $N$ permeable particles bouncing on a static mirror in the 1D space in the presence of gravitational field can be written in the form 
\be
H_0=\sum_{i=1}^N\left(\frac{p_i^2}{2}+x_i\right), \quad x_i\ge 0,
\ee 
where we assume that the mirror is located at $x_i=0$, the particles are of unit masses, and gravitational acceleration is equal to one. Energies $E_i$ of the particles are constants of motion, the system is separable, and in the classical description it is convenient to introduce a canonical transformation to action-angle variables  $(I_i,\theta_i)$, where for each particle~\cite{Lichtenberg1992,Flatte1996,Buchleitner2002}
\be
  I_{i}=\frac{\left(2 E_{i} \right)^{3/2}}{3\pi}, \quad \theta_{i}=\pi -\frac{\pi}{\sqrt{2E_{i}}}p_{i}.
\label{a-a_per}  
\ee
Then, the Hamiltonian takes the form
\be
H_0=\sum_{i=1}^N \frac{(3\pi I_{i})^{2/3}}{2},
\label{haa}
\ee
and solutions of the Hamilton's equations read \be
I_i(t)={\rm constant}, \quad \theta_i(t)=\Omega(I_i)\;t+\theta_i(0) \quad (\text{mod } 2\pi), 
\label{sol1}
\ee
where 
\be
\Omega(I_i)=\frac{\partial H_0}{\partial I_i}=\left(\frac{\pi^2}{3I_i}\right)^{1/3},
\label{Omegas}
\ee
is the period of motion of the $i$-th particle. Equations~\eqref{sol1} show that $N$ permeable particles move on an $N$-torus in $2N$-dimensional phase space. The explicit form of the canonical transformation between the action-angle variables and the Cartesian variables of the permeable particles is 
\be
x_i=\frac12\left(\frac{3I_{i}}{\pi^2}\right)^{2/3}(2\pi -\theta_{i})\theta_{i},\quad
p_{i}= \left(\frac{3 I_{i}}{\pi^2}\right)^{1/3}(\pi-\theta_{i}).
\label{trans_explicite}
\ee 
Note that each pair of the action-angle variables $(I_i,\theta_i)$ is always associated with one particle only.

\subsection{Action-angle variables for impenetrable particles with different energies}
\label{aaimpen}

The system of $N$ impenetrable point-like particles (i.e. hard balls with zero radius) of unit masses bouncing on a static mirror is integrable~\cite{wojtkowski1990}. Suppose that the particle positions are labelled in ascending order, i.e. $0\le x_1\le x_2\le \hdots\le x_i<\hdots\le x_N$. All scattering events are elastic. The lowermost particle bounces on the mirror at $x_1=0$ reversing its momentum $p_1 \to -p_1$. When any two particles collide at $x_i=x_{i+1}$, they exchange their momenta $p_i \leftrightarrows p_{i+1}$ which also means that their energies are swapped $E_i \leftrightarrows E_{i+1}$. Between the collisions, the particles do not change their energies and they follow a set of trajectories similar to those in the case of the permeable particles. After a collision, the set of trajectories remains the same but some particles exchange their paths.

One can integrate the Hamilton's equations of motion using the Cartesian variables with an additional condition that every time two particles collide at some position $x_i=x_{i+1}$ their momenta have to be swapped. However, if all particle energies are different, we can define novel action-angle variables $({\cal I}_{i'},\vartheta_{i'})$ where the index $i'$ is not associated with the number of a particle but rather with a given value of energy. Suppose that initially we have chosen a set of particle energies $\{E_i\}$ that are all different and in ascending order form a sequence ${\cal E}_1<{\cal E}_2<\hdots<{\cal E}_{i'}<\hdots<{\cal E}_N$. It is important to note that after sorting this set, the element ${\cal E}_{i'}$ is in general  different from $E_{i'}$. Then, we can define the canonical transformation to the action-angle variables in the following manner [cf.~\eqref{a-a_per}]
\bea
{\cal I}_{i'}&=&\frac{(2{\cal E}_{i'})^{3/2}}{3\pi}, \\
\vartheta_{i'}&=&\pi-\frac{\pi}{\sqrt{2 {\cal E}_{i'}}} \sum_{i=1}^N p_i\, \delta_{E_i,{\cal E}_{i'}}, \label{ftran}
\eea
where $\delta_{E_i,{\cal E}_{i'}}=1$ for $E_i={\cal E}_{i'}$ and zero otherwise. The Hamiltonian of the system has the same form as~\eqref{haa} and integrating the corresponding Hamilton's equations yields the following results, 
\be
{\cal I}_{i'}(t)={\rm constant}, \quad \vartheta_{i'}(t)=\Omega({\cal I}_{i'})\;t+\vartheta_{i'}(0) \quad (\text{mod }2\pi).
\label{sol2}
\ee 
Note that in the present case of impenetrable particles with different energies, the phase space is restricted to ${\cal I}_1< {\cal I}_2<\hdots< {\cal I}_{i'}<\hdots< {\cal I}_N$.

Having obtained solutions~\eqref{sol2} of the equations of motion in the action-angle variables, one can transform back to the Cartesian variables using the inverse canonical transformation. To do that, we calculate
\be
{\cal X}_{i'}=\frac12\left(\frac{3 {\cal I}_{i'}}{\pi^2}\right)^{2/3}(2\pi -\vartheta_{i'})\vartheta_{i'},\quad
{\cal P}_{i'}= \left(\frac{3 {\cal I}_{i'}}{\pi^2}\right)^{1/3}(\pi-\vartheta_{i'}).
\label{inverse_tilde}
\ee 
Next, as the particles are impenetrable (i.e. $0\le x_1\le x_2\le\hdots\le x_N$) in order to associate the pairs $({\cal X}_{i'},{\cal P}_{i'})$ with the actual positions and momenta of the particles one has to put the set $\{{\cal X}_{i'}\}$ in ascending order and then associate the corresponding pairs $x_i={\cal X}_{i'}$ and $p_i={\cal P}_{i'}$.

The solutions~\eqref{sol2} indicate that in the action-angle variables $({\cal I}_{i'},\vartheta_{i'})$, the evolution of the system takes place on an $N$-torus in the $2N$-dimensional phase space. However, for different segments of the trajectory on an $N$-torus single angle $\vartheta_{i'}$ is associated with different particles because the action-angle variables $({\cal I}_{i'},\vartheta_{i'})$ do not correspond to the particles themselves but rather to the energies ${\cal E}_{i'}$.

\subsection{Two impenetrable particles with equal energies} 
The problem of $N$ impenetrable particles with all equal energies has to be treated using the action-angle variables $(I_i,\theta_i)$ previously introduced for permeable particle (see \sref{aaper}) rather than using the novel action-angle variables $({\cal I}_{i'},\vartheta_{i'})$ because the transformation \eqref{ftran} is not defined in this case. Note that these variables $(I_i,\theta_i)$ are associated with particles and not with energies unlike the variables $({\cal I}_{i'},\vartheta_{i'})$ described in \sref{aaimpen}. 

It is now instructive to summarise the results of reference~\cite{Giergiel2021} where the case of a particle moving in a 2D space between two mirrors forming a wedge of $\pi/4$ was considered. As we have mentioned in \sref{intro}, such a system maps onto a 1D problem of two impenetrable particles with equal energies, i.e.  $N=2$, $0\le x_1 \le x_2$, and $I\equiv I_1=I_2$, bouncing on a single mirror. The motion of the system in the $(\theta_1,\theta_2)$ space was shown to take place on a M\"obius strip.

\begin{figure}[ht!]
   \centering
   \includegraphics[]{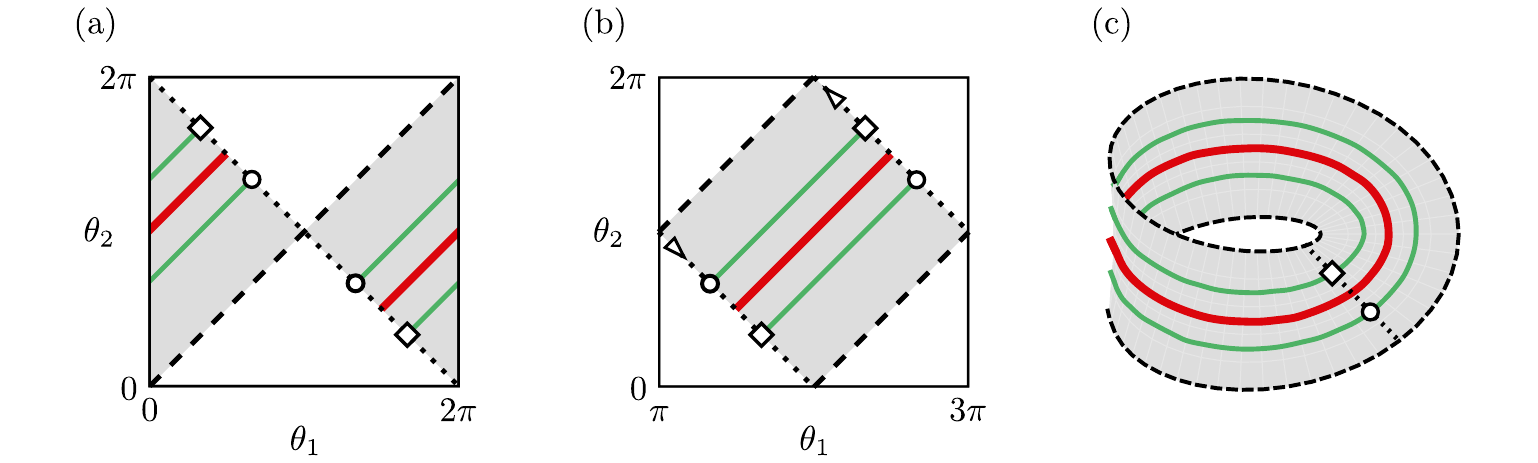}
    \caption{Two impenetrable particles with equal energies move on the M\"obius strip. (a) The motion of the system in $(\theta_1,\theta_2)$ space is restricted by the condition $(2 \pi-\theta_1)\, \theta_1\le (2\pi-\theta_2)\,\theta_2 $ (grey area). The nontrivial boundaries of this area correspond to two solutions of equation \eqref{x1x2col}. The first solution $\theta_1=\theta_2$ forms the edge of the strip (dashed black line), and the second solution $\theta_2 = 2\pi-\theta_1$ twists one of the ends of that strip (dotted black line). The latter implies the M\"obius strip geometry, which is easier to visualise after connecting the trivial edges $\theta_1=0$ and $\theta_1=2\pi$ by shifting the left grey triangle to the right (b) and then gluing two dotted lines so that the arrows match up~(c). The solid green lines are the segments of a typical trajectory on this manifold. The entire trajectory is obtained by connecting the circles (squares) to each other. Among such trajectories there is a unique one with period twice shorter than the others (thick red line).}
    \label{fig:Mobius}
\end{figure}

This manifold can be identified in the following way. In the case of the same particle energies, the condition $0 \le x_1\le x_2$ reduces to $0\le(2 \pi-\theta_1)\, \theta_1\le (2\pi-\theta_2)\,\theta_2 $ meaning that the motion of the system is restricted to half of the space explored by the permeable particles, see \fref{fig:Mobius}a.
When the impenetrable particles collide, i.e. $x_1=x_2$ or equivalently 
\be
(2\pi-\theta_1)\theta_1=(2\pi-\theta_2)\theta_2,
\label{x1x2col}
\ee 
their momenta have to be swapped, i.e. $p_1\rightleftarrows p_2$ or equivalently $\theta_1\rightleftarrows \theta_2$. Equation~\eqref{x1x2col} has a solution  $\theta_1=\theta_2$ corresponding to two impenetrable particles moving together along identical trajectories with the same momenta $p_1=p_2$. This trajectory forms an edge of a strip in the $(\theta_1,\theta_2)$ space. However, there is another solution with $\theta_2=2\pi-\theta_1$. In that case swapping the momenta corresponds to twisting one of the ends of that strip and joining them together to form the M\"obius strip as illustrated in \fref{fig:Mobius}(a)-(c).

\subsection{Three impenetrable particles with equal energies} \label{sec3}

Here, we extend the analysis from the previous section to the case of $N=3$ impenetrable particles with equal energies using the action-angle variables $(I_i,\theta_i)$ introduced for permeable particles in \sref{aaper}. The motion of the system in $(\theta_1,\theta_2,\theta_3)$ space is restricted by the condition $0 \le (2 \pi-\theta_1) \theta_1\le (2\pi-\theta_2)\theta_2 \le  (2\pi-\theta_3)\theta_3$ for $I \equiv I_1=I_2=I_3$. This corresponds to $1/6$ of the space explored by permeable particles with nontrivial boundaries given by the solutions of the following equations
\begin{align}
\begin{split}    
    \label{N3collisions}
    (2\pi-\theta_1)\theta_1 &= (2\pi-\theta_2)\theta_2, \\
    (2\pi-\theta_2)\theta_2 &= (2\pi-\theta_3)\theta_3, \\
    (2\pi-\theta_1)\theta_1&=(2\pi-\theta_2)\theta_2 =(2\pi-\theta_3)\theta_3,
\end{split}
\end{align}
describing three possible particle collisions.

\begin{figure}[ht!]
    \centering
    \includegraphics[]{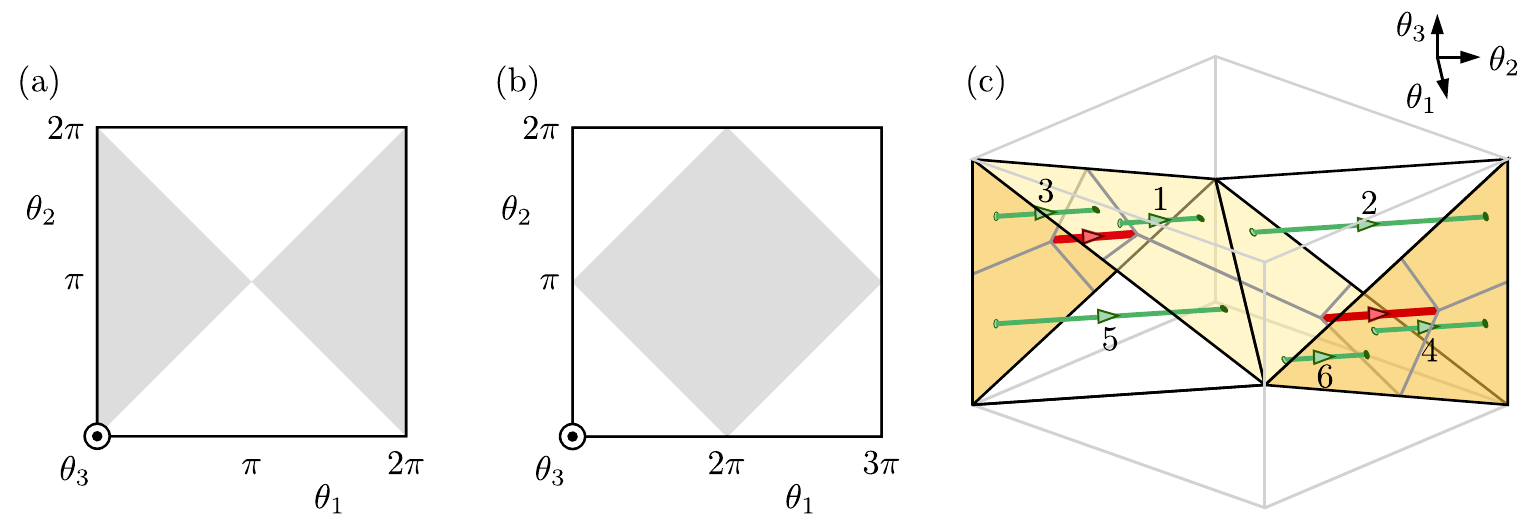}
    \caption{The motion of three impenetrable particles with equal energies in $(\theta_1,\theta_2,\theta_3)$ space is restricted by the condition $0 \le (2 \pi-\theta_1) \theta_1\le (2\pi-\theta_2)\theta_2 \le  (2\pi-\theta_3)\theta_3$. (a)-(b) To simplify the further identification of this manifold, we shift the cuboid $(\theta_1\in (0,\pi]))\times(\theta_2,\,\theta_3\in(0,2\pi))$ by $2\pi$
    in accordance with $2\pi$-periodicity of $\theta_1$. For clarity, the projection of the restricted space onto the $(\theta_1,\theta_2)$ plane is shown before (a), and after (b) the shift. (c) As a result, $(\theta_1,\theta_2,\theta_3)$ space explored by the system consists of two tetrahedrons joined by the common edge. Their faces correspond to nontrivial boundary conditions described by the solutions of equations~\eqref{N3collisions}. 
    The surface of the manifold the particles move on will be formed by the four uncoloured triangles, while gluing the remaining yellow and orange ones colour-wise provides continuity of the motion inside the manifold, i.e. making the typical solid green trajectory continuous. The motion along the green trajectory takes place according to the labels given to the trajectory segments (1–6) --- each segment corresponds to the motion of the particles between collisions. Collisions between the first particle and the second lowermost particle correspond to the orange area and between the second and third ones to the yellow area. Among such trajectories there is a unique (central) one with a period thrice shorter than the others (thick red line). It runs through the orthocentre of the triangles.}
 \label{fig:N3impe}
\end{figure}

To facilitate the identification of the manifold the particles move on, we get rid of the trivial boundary conditions in accordance with $2\pi$-periodicity of $\theta_1$. As a result, the restricted $(\theta_1,\theta_2,\theta_3)$ space consists of two tetrahedrons joined by a common edge. Their faces correspond to nontrivial boundary conditions described by the solutions of equations \eqref{N3collisions}, see \fref{fig:N3impe}(a)-(c) for details.
After connecting the faces of tetrahedrons, we observe that the manifold depicted in \fref{fig:N3impe}(c) is homeomorphic to a solid torus, i.e. the product of a circle and a 2-disc. 

\begin{figure}[ht!]
    \centering
    \includegraphics[]{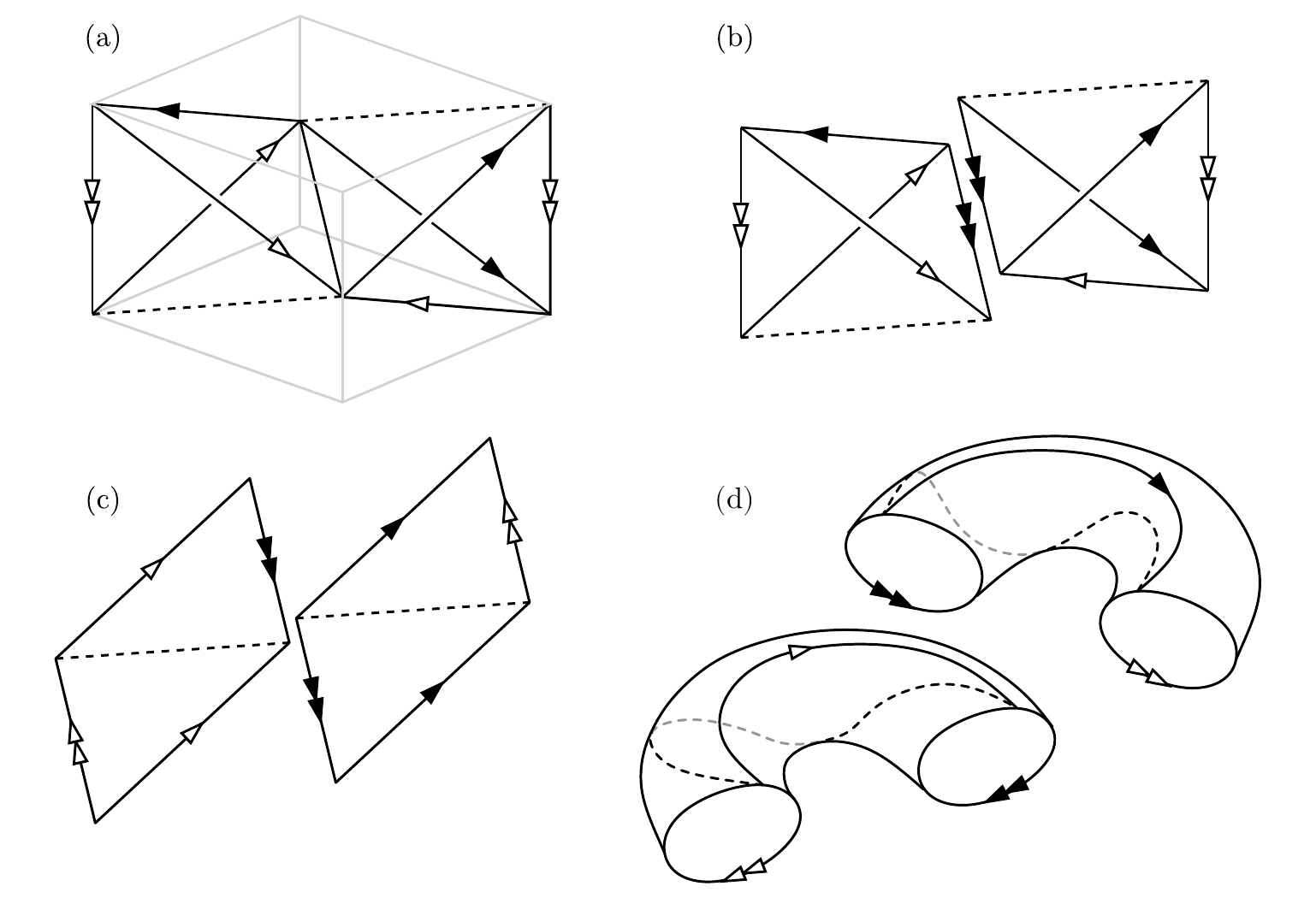}
    \caption{The boundary of the manifold depicted in \fref{fig:N3impe}(c) is homeomorphic to a 2-torus. (a) By analysing the motion of the system, we determine how the edges of tetrahedrons should be glued together. That is, the corresponding arrows have to match up. (b) To visualise this process, we separate the tetrahedrons by adding one more edge identification (lines with double-headed black arrows). Then, we focus on the boundary triangles only (c), i.e. those uncoloured in \fref{fig:N3impe}(c), and obtain two halves of a 2-torus by gluing the corresponding single-headed arrows (d). Finally, the appropriate connection of the double-headed arrows leads to the 2-torus.}
    \label{fig:N3boundary}
\end{figure}

Since the models do not yield a ready visualisation of that manifold in the 3D space, we must proceed as follows. First, we determine its boundary formed by the four uncoloured faces of the tetrahedrons in \fref{fig:N3impe}(c). By analysing the motion of the system, we identify the edges of tetrahedrons, and then glue them together to obtain a 2-torus $\mathbb{S}^1\times\mathbb{S}^1$, see \fref{fig:N3boundary}(a)-(d).
    
Second, we determine the geometry of tetrahedron interiors. It can be written as the product of the segment $[0,1]$ and the interior of the 2-simplex $int \Delta^2$, where $\{1\}\times int \Delta^2$ and $\{0\}\times int \Delta^2$ are identified by a $\frac{2\pi}{3}$ clockwise rotation. This rotation ensures the continuity of the system trajectories. As a result, we analyse a fibre bundle of open 2-discs over the unique orbit, i.e. thick red trajectory in \fref{fig:N3impe}(c). Its neighbourhood can be taken arbitrarily large to completely fill the interior of the manifold. Therefore, it is homeomorphic to the interior of a solid torus. Note that we employed the fact that the open 2-disc and interior of the 2-simplex are homeomorphic.
    
In accordance with the Loop Theorem about 3-manifolds~\cite{Hempel1976} if both the interior and the boundary of a manifold are that of a solid torus, the manifold is homeomorphic to a solid torus itself. We refer to~\cite{Hatcher2007}, Proposition 3.4 for the complete argument.

We note that contrary to the $N=2$ case, we obtained an orientable manifold. Perhaps it is not surprising. The M\"obius strip is a non-orientable fibre bundle over a circle, and it is well-known that a topological fibred product of two M\"obius strips is homeomorphic to the orientable solid torus. We suspect that this phenomenon will persist for $N>3$, where the phase space for the particles will prove to be a fibred product of $N-1$ M\"obius strips, but this does not follow immediately from the analysis of the model for $N=3$, and generalisations do not seem straightforward.

\section{Periodically driven system}
\label{two}

In the present section, we turn on the mirror oscillations in order to analyse the resonant motion of $N$ impenetrable particles of unit masses, where the lowermost particle is bouncing on this oscillating mirror in the presence of gravitational field. The Hamiltonian of the system reads
\be
H=\sum_{i=1}^N\left(\frac{p_i^2}{2}+x_i\right)+F[x_1-f(t)], 
\label{hlab1}
\ee
where $x_1\le \hdots \le x_N$ and $F(x)$ is the reflecting potential of the mirror oscillating with the frequency $\omega$, i.e. the position of the mirror is described by $f(t+2\pi/\omega)=f(t)$. It is convenient to switch to the reference frame of the oscillating mirror (in the following, we will refer to this frame as the laboratory frame) by means of the following canonical transformation
\be
\tilde x_i=x_i-f(t), \quad \tilde p_i=p_i-f'(t),
\ee
which leads to 
\be
H=\sum_{i=1}^N\left(\frac{p_i^2}{2}+x_i+f''(t)x_i\right)+F(x_1),
\label{hamosc}
\ee
where the tilde marks have been omitted. In the further analysis, we drop the $F(x_1)$ term as the hard wall potential of the mirror at $x_1=0$ is assumed, keeping in mind that motion of the particles takes place for $0 \le x_1 \le x_2 \le \hdots \le x_N$. In the next subsections, we derive the effective secular Hamiltonian describing the dynamics of $N$ impenetrable particles close to a resonant trajectory separately for particles with all equal and all different energies.

\subsection{Resonant motion of $N$ impenetrable particles with different energies}
\label{resonant_different_energies}

For different energies of $N$ impenetrable particles the action-angle variables $({\cal I}_{j},\vartheta_j)$ introduced in \sref{aaimpen} are used to rewrite the Hamiltonian~\eqref{hamosc} in the following way
\be
H=\sum_{j=1}^N\left[ \frac{(3\pi{\cal I}_{j})^{2/3}}{2}+f''(t)\frac12\left(\frac{3 {\cal I}_{j}}{\pi^2}\right)^{2/3}(2\pi-\vartheta_{j})\vartheta_{j}\right],
\label{hamimpdiffen}
\ee
where $j$ labels particle energies in the ascending order, ${\cal E}_1<{\cal E}_2<\hdots<{\cal E}_N$ or equivalently ${\cal I}_1<{\cal I}_2<\hdots<{\cal I}_N$. Then, the time periodic function $f''(t)$ is expanded into a Fourier series
\be
f''(t)=\lambda \sum_{k\ne 0} f_k e^{ik\omega t}, \label{driving}
\ee
where the parameter $\lambda$ controls how strongly the particles are driven by the oscillating mirror. We consider the case of a resonant motion given by the condition ${\cal I}_{j}\approx {\cal I}_{j,0}$, where ${\cal I}_{j,0}$ are the resonant values of the actions in the static mirror case, i.e. $\lambda=0$. The resonant values of the action ${\cal I}_{j,0}$ satisfy
\be
\omega=s_1\Omega({\cal I}_{1,0})=s_2\Omega({\cal I}_{2,0})=\hdots =s_N\Omega({\cal I}_{N,0}),
\label{res_cond}
\ee
with integer $s_1<s_2<\hdots<s_N$. To obtain an effective Hamiltonian describing the motion of particles close to a resonant trajectory, we apply the classical secular approximation~\cite{Lichtenberg1992}. First, we switch from the laboratory frame to the reference frame moving along a resonant trajectory,
\be
\tilde\vartheta_{j}=\vartheta_{j}-\frac{\omega}{s_{j}}t, \quad \tilde {\cal I}_{j}={\cal I}_{j},
\ee
and obtain an intermediate form of the Hamiltonian
\be
H=\sum_{j=1}^N\left[\frac{(3\pi\tilde{\cal I}_{j})^{2/3}}{2}-\frac{\omega}{s_j}\tilde {\cal I}_j+\lambda\sum_{k\ne 0}f_ke^{ik\omega t}\sum_{n_j}h_{n_j}(\tilde{\cal I}_j)e^{in_j(\tilde\vartheta_j+\omega t/s_j)}\right],
\label{hmov}
\ee
where
\be
h_{n_j}(\tilde{\cal I}_j)=-\frac{1}{n_j^2}\left(\frac{3 \tilde{\cal I}_{j}}{\pi^2}\right)^{2/3},
\ee
 for $n_j\ne 0$, and $h_{0}(\tilde{\cal I}_j)=(\pi^2 \tilde{\cal I}_j^2/3)^{1/3}$.
Then, we average the Hamiltonian~\eqref{hmov} over time keeping all dynamical variables ${\tilde{\cal{I}}}_j$ and $\tilde{\vartheta}_j$ fixed because close to the resonant trajectory, i.e. for $\tilde{\cal I}_{j}\approx{\cal I}_{j,0}$, they vary slowly if the driving strength $\lambda$ is small enough~\cite{Lichtenberg1992}. It results in the effective secular Hamiltonian 
\bea
H\approx \sum_{j=1}^N\left(\frac{P_j^2}{2m_j} +\lambda\sum_{n_j}f_{n_j}h_{n_js_j}({\cal I}_{j,0})e^{in_js_j\tilde\vartheta_j}\right),
\label{heff}
\eea
where a constant term has been omitted and the effective masses are
\be
m_j=-\frac{1}{2}\left(\frac{9\, {\cal I}^2_{j,0}}{\pi}\right)^{2/3}.
\ee 
Obtaining Hamiltonian~\eqref{heff} includes performing Taylor series expansions around the resonant values of actions ${\cal I}_{j,0}$
\be
\frac{(3\pi\tilde{\cal I}_{j})^{2/3}}{2}-\frac{\omega}{s_j}\tilde {\cal I}_j\approx {\rm constant}+\frac{P_j^2}{2m_j},
\label{taylor}
\ee 
where $P_j=\tilde{\cal I}_{j}-{\cal I}_{j,0}$. The linear term in~\eqref{taylor} vanishes due to the resonance condition~\eqref{res_cond}.

The effective Hamiltonian~\eqref{heff} can be interpreted not only as describing a system of $N$ impenetrable particles but also as the Hamiltonian of a single fictitious {\it particle} moving in an $N$-dimensional space in the presence of a time-independent potential. The potential is separable and its shape depends on the way the mirror is driven. The choice of the Fourier components $f_k$ of the periodic mirror oscillations shapes the effective potential. For example, in the case of a simple harmonic driving $f''(t)=\lambda\cos(\omega t)$ we obtain 
\be
H\approx \sum_{j=1}^N\left[\frac{P_j^2}{2m_j} +\lambda h_{s_j}({\cal I}_{j,0})\cos(s_j\tilde\vartheta_j)\right],
\label{heff1}
\ee
meaning that the system behaves like a {\it particle} in $N$-dimensional space in the presence of $s_j$ potential wells along the $j$-th direction.

In general, the motion of a single particle bouncing on an oscillating mirror in a 1D space is regular provided the mirror amplitude $\lambda$ is not too large. Especially, the resonant trajectories surrounded by KAM tori~\cite{Lichtenberg1992} can be observed. This picture breaks down if the amplitude of the mirror oscillations is greater than $\lambda \approx$ 0.2~\cite{Giergiel2020} according to the Chirikov criterion~\cite{Chirikov1979}. The $N$-particle system we consider here separates into $N$ independent motions along 1D spaces, cf.~equation~\eqref{hamimpdiffen}. Therefore, the analysis of chaos in the $N$-particle system reduces to the analysis of chaos in the 1D cases. As we restrict ourselves to the driving amplitude $\lambda$ small enough, the motion of the system under consideration is regular.

The system is suitable for investigation of classical synchronisation phenomena. 
If $N$ classical particles are replaced by $N$ impenetrable clouds formed by classical particles, and the particles within each cloud can interact weakly, one can look for the range of parameters where synchronous evolution of the system with a period different than the driving period is observed. This kind of phenomenon is related to the so-called classical discrete time crystals \cite{Shapere2012}.

Having arrived at the Hamiltonian~\eqref{heff1}, it is possible to move to the quantum description with all $s_j\gg 1$, where various $N$-dimensional condensed matter phenomena can be investigated in the system at hand. For example, proper choice of the Fourier components $f_k$~\eqref{hmov} allows one to introduce disorder in the crystalline structure in~\eqref{heff1} or create a crystal with nontrivial topological properties~\cite{SachaTC2020}. Quantum analysis of these phenomena is left for future work.

\subsection{Resonant motion of $N=3$ impenetrable particles with equal energies}

In \sref{sec3} we analysed $N=3$ impenetrable particles with equal energies that periodically bounce on the static mirror. When the mirror starts oscillating and drives the particles resonantly, we can derive an effective secular Hamiltonian similarly as in \sref{resonant_different_energies}. In order to do that we cannot use the same action-angle variables $({\cal I}_j,\vartheta_j)$ as in \sref{resonant_different_energies} because they are not well defined in this case. However, the effective secular Hamiltonian can be obtained with the help of the action-angle variables $(I_j,\theta_j)$ derived for permeable particles, see \sref{aaper}. In terms of these variables, the condition $0\le x_1\le x_2\le x_3$ fulfilled by $N=3$ impenetrable particles reads
\be
0 \le I_1^{2/3}(2 \pi-\theta_1) \theta_1\le I_2^{2/3}(2\pi-\theta_2)\theta_2 \le I_3^{2/3}(2\pi-\theta_3)\theta_3.
\label{inequalities}
\ee
When two particles collide, i.e. $x_j=x_{j+1}$, they exchange their momenta $p_j\leftrightarrows p_{j+1}$ which means that both their actions $I_j\leftrightarrows I_{j+1}$ and angles $\theta_j\leftrightarrows \theta_{j+1}$ have to be swapped. As a result, the Hamiltonian of the system, 
\be
H=\sum_{j=1}^3 \frac{(3\pi I_j)^{2/3}}{2} + f''(t)\frac12\sum_{j=1}^3 \left(\frac{3 I_{j}}{\pi^2}\right)^{2/3}(2\pi-\theta_{j})\theta_{j}, 
\label{hdriv}
\ee
is invariant under such a transformation. 

Along the resonant trajectory all particles have the same energies, i.e. the same value of the action $I_0$ that fulfils the condition $\omega=s \Omega(I_0)$ where $s$ is an integer. To obtain the effective Hamiltonian within the classical secular approximation, we first switch from the laboratory frame to the reference frame moving along a resonant trajectory, i.e. $\tilde\theta_j=\theta_j-\omega t/s$ and $\tilde I_j=I_j$. Then, we perform averaging of the resulting Hamiltonian over time keeping $\tilde I_j$ and $\tilde\theta_j$ fixed because they vary slowly in the vicinity of the resonant trajectory provided the perturbation is weak. Note that even though in the moving reference frame $\tilde\theta_j$ are kept fixed, in the laboratory frame the variables $\theta_j=\tilde\theta_j+\omega t/s$ evolve with time $t$ and for certain points in time  collisions of the particles take place. As a result, one has to swap the values of $\tilde\theta_j$ and $\tilde I_j$ of the colliding particles. However, it has no effect on the Hamiltonian because it is invariant under exchanges $\tilde I_j\leftrightarrows \tilde I_{j+1}$ and $\tilde\theta_j\leftrightarrows \tilde\theta_{j+1}$. Consequently, we obtain an effective secular Hamiltonian very similar to~\eqref{heff} or~\eqref{heff1} -- for exemplary driving $f''(t)=\lambda\cos(\omega t)$, one gets
\be
H\approx \sum_{j=1}^3\left[\frac{P_j^2}{2m} +\lambda h_{s}(I_0)\cos(s\tilde\theta_j)\right],
\label{heff1equal}
\ee
where $P_j=\tilde I_j-I_0$ and the effective masses $m=-\left(9\, {\cal I}^2_{j,0}/\pi\right)^{2/3}/2$ are all equal. 

As a consequence of using action-angle variables $(I_j,\theta_j)$ for permeable particles to derive the effective secular Hamiltonian of $N$ impenetrable particles with all equal energies, one has to monitor the evolution of those variables in the laboratory frame, i.e. $I_j(t)=\tilde I_j(t)$ and $\theta_j(t)=\tilde\theta_j(t)+\omega t/s$, during the integration of the Hamilton's equations generated by~\eqref{heff1equal}. Then, if any of the inequalities in~\eqref{inequalities} is saturated, the values of the corresponding angles $\tilde\theta_j \leftrightarrows \tilde\theta_{j+1}$ and actions $\tilde I_j \leftrightarrows \tilde I_{j+1}$ have to be swapped. In contrary, the dynamics of $N$ particles with different energies generated by Hamiltonian~\eqref{heff} or~\eqref{heff1} is much simpler (see \sref{resonant_different_energies}). With the help of the action-angle variables $(\tilde{\cal I}_j,\tilde{\vartheta}_j)$ we do not have to monitor particle collisions because they are associated with given values of energies rather than with the particles themselves as is the case for $(I_j,\theta_j)$ variables, see \sref{aaimpen}.

\section{Summary and outlook}
In the first part of the paper, we considered a 1D problem of $N$ impenetrable particles over a static mirror in the presence of gravitational field. We analysed the cases of all different and all equal particle energies. For the case of different energies we introduced novel action-angle variables $({\cal I}_i,{\vartheta}_i)$ which label not particles themselves, but their energies instead. Such an approach allowed us to bypass the issue of complicated boundary conditions due to particle collisions. We readily identified that the evolution of the system takes place on an $N$-torus in $2N$-dimensional phase space. For the case of equal energies, we resolved to using standard action-angle variables $(I_i,\theta_i)$~\cite{Lichtenberg1992} which force the cumbersome analysis of the boundary conditions. Taking the problem of $N=2$ particles as the baseline~\cite{Giergiel2021} we investigated the problem of $N=3$ to find that the evolution takes place on a solid torus.

We carry on the distinction between the cases of all equal and all different particle energies by employing both types of action-angle variables to derive an effective secular Hamiltonian for resonant motion of particles when the mirror oscillates periodically in time. This effective time-independent Hamiltonian is obtained in the reference frame moving along a resonant trajectory. The effective Hamiltonian is additionally interpreted as describing a fictitious \textit{particle} in an $N$-dimensional effective potential. A proper choice of the Fourier components of the mirror oscillations allows us to shape the effective potential forming various crystalline structures. This result opens up a venue for investigations of variety of $N$-dimensional condensed matter phenomena.

Experimentally, the system under consideration can be realised by setting up $N$ clouds of strongly repulsive atoms bouncing on top of each other over an oscillating mirror in the 1D space~\cite{Giergiel2018a,Giergiel2020}. For $N=2$ or $N=3$, there is also another way to realise the system. Taking a single atomic cloud bouncing between the walls of the 2D or 3D wedge formed by properly arranged atom mirrors is equivalent to the system investigated here. 

The results presented here form a basis for the investigation of quantum time crystals. Note that although, we identified the crystalline structures in the reference frame moving along a resonant trajectory using a time-independent effective Hamiltonian, they will be observed in the time domain when we return to the laboratory frame~\cite{SachaTC2020}. Our results open novel directions for the realisation of quantum time crystals and investigation of condensed matter phenomena in the time domain. 

\section*{Acknowledgements}

We thank Czcibor Ciostek for the fruitful discussions and for a critical reading of this paper. 
We are grateful to Peter Hannaford for reading the manuscript and for suggesting improvements. This work was supported by the National Science Centre, Poland via Projects   No.  2018/31/B/ST2/00349 (W.G. and K.S.), QuantERA Programme No. 2017/25/Z/ST2/03027 (A.K.).

\section*{References}
\bibliographystyle{iopart-num}
\bibliography{references.bib}

\end{document}